\documentclass[epj]{svjour}
\usepackage{graphics}
\begin{document}
\title{Search for beta-delayed proton emission from $^{11}$Be}
\author{ K. Riisager\inst{1}\thanks{email:kvr@phys.au.dk}
 \and M.J.G. Borge\inst{2,3} \and J.A. Briz\inst{3} 
 \and M. Carmona-Gallardo\inst{4} 
 \and O. Forstner\inst{5} \and L.M. Fraile\inst{4} \and H.O.U. Fynbo\inst{1}
 \and A. Garzon Camacho\inst{3}
 \and J.G. Johansen\inst{1} \and B. Jonson\inst{6}
 \and M.V. Lund\inst{1} \and J. Lachner\inst{5}
 \and M. Madurga\inst{2} \and S. Merchel\inst{7} \and E. Nacher\inst{3}
 \and T. Nilsson\inst{6} \and P. Steier\inst{5}
 \and O. Tengblad\inst{3} \and V. Vedia\inst{4}
}                     
%
%
\institute{Department of Physics and Astronomy, Aarhus University,
 DK--8000, Aarhus C, Denmark
  \and ISOLDE, EP Department, CERN, CH--1211 Geneve 23, Switzerland
  \and Instituto de Estructura de la Materia, CSIC, E--28006 Madrid, Spain
  \and Grupo de F\'{\i}sica Nuclear and IPARCOS, Universidad Complutense de Madrid,
CEI Moncloa, E-28040 Madrid, Spain
  \and Faculty of Physics, University of Vienna, W\"{a}hringer Strasse
  17, A--1090 Wien, Austria
  \and Institutionen f\"{o}r Fysik, Chalmers Tekniska H\"{o}gskola, SE--41296
  G\"{o}teborg, Sweden
  \and Helmholtz-Zentrum Dresden-Rossendorf, D--01328 Dresden, Germany
 }
\date{Received: date / Revised version: date}
%
\abstract{We report on an attempt to reproduce the observation of
  $\beta^-$-delayed proton emission from $^{11}$Be through detection
  of the final state nucleus $^{10}$Be with accelerator mass
  spectrometry. Twelve samples were collected at the ISOLDE facility
  at CERN at different separator settings, allowing tests of different
  sources of contamination to be carried out. The observed amounts of
  $^{10}$Be per collected $^{11}$Be rule out several contamination
  sources, but do not agree internally. Formation of BeH molecular
  ions in the ion source may explain our data, in which case an upper
  limit of the $\beta$p branching ratio of $2.2 \cdot 10^{-6}$ can be derived.
%
\PACS{
      {23.40.-s}{$\beta$ decay}   \and
      {27.20.+n}{$6 \leq$ A $\leq 19$}
     } 
} 
\maketitle
\section{Introduction}
\label{sec:intro}
Beta-decay has long been recognized as a powerful probe of nuclear
structure. It gains further in versatility when moving to nuclei
far from the line of beta-stability as beta-delayed particle
emission become energetically possible, see \cite{Pfu12,Bla08} for general
reviews.  Certain beta-delayed processes turn out \cite{Jon01} to
occur only in near-dripline nuclei, the one we focus on here is
beta-minus delayed proton emission. This process is only allowed for a
few neutron-rich nuclei and was predicted \cite{Bay11} to have a very
small branching ratio even for the most promising case, that of
$^{11}$Be. The first experiments \cite{Ang98,Bor13,Rii14} to look for this
decay focussed on detecting the final nucleus $^{10}$Be rather than
the emitted proton.  In the latest of these \cite{Rii14} evidence for
the decay was found with an intensity $8.3(9) \cdot 10^{-6}$, orders of
magnitude above the preceeding theoretical prediction of
$3.0 \cdot 10^{-8}$ \cite{Bay11}. Apart from the intrinsic interest of
this result, recent suggestions of alternative neutron decay branches
\cite{For18,Pfu18} have added motivation for the study of this
particular decay. It is therefore important to have a careful
assessment of possible systematic errors that may influence the
obtained results. This paper describes a more extensive series of
tests carried out to check the validity of the reported branching
ratio.

An overview of the current state of knowledge on the beta decay of
$^{11}$Be was given in \cite{Bor13}. A recent experimental result 
\cite{Ref18} clarified the situation for the beta-delayed
$\alpha$-decay branch of $^{11}$Be.

\section{Experimental procedures}
\label{sec:exp}

\begin{figure}
  \resizebox{0.48\textwidth}{!}{%
    \includegraphics{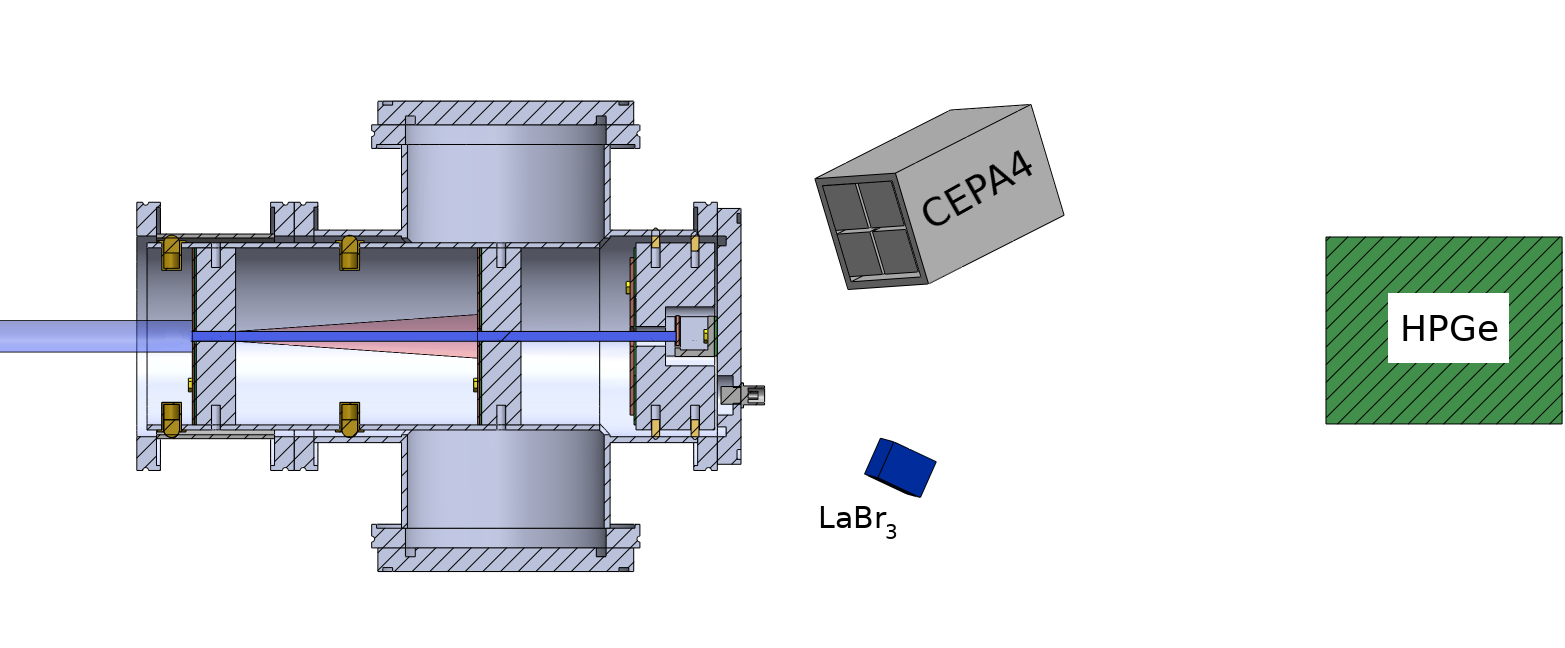}
  }
\caption{The set-up: the $^{11}$Be beam is coming from the left,
  passes several collimators and is collected on the Cu sample that
  is mounted on the end flange. The on-line gamma activity is recorded
  in a HPGe at zero degrees (far right) and in LaBr-detectors
  positioned at both sides. See the text for details.}
\label{fig:setup} 
\end{figure}

The basic idea is as in the earlier experiments \cite{Bor13,Rii14} to
collect $^{11}$Be samples, determine their intensity on-line via the
$\gamma$-decays and measure the amount of produced
$^{10}$Be via a subsequent accelerator mass spectrometry (AMS) step.
Samples collected at different settings of the ISOLDE mass separator
allow checking the reproducibility of the result as well as testing
for different sources of contamination.
Figure \ref{fig:setup} gives an overview of the experimental set-up at
the source position.

The distance from the collection point to the outer side of the flange
of the collection chamber is 31 mm.
A HPGe detector was positioned 632(2) mm downstream from the flange
in continuation of the beam line direction.
A LaBr$_3$(Ce) and a CEPA4 (combined LaBr$_3$(Ce) and LaCl$_3$(Ce)
\cite{Nac15}) detector were 
placed at either side of the chamber in the downstream direction in order
not to block the line-of-sight to the HPGe detector, the distances from the flange of the
chamber to the front edge of the LaBr$_3$ and the CEPA4 were 88 mm and
126 mm, respectively, corresponding to angles of 40--50 degrees from
the beam line direction.
The cylindrical LaBr$_3$ detector had a radius of 10 mm and a length of 20 mm, while
the CEPA4 detector was formed by four squared crystals optically
isolated, each of side length 27 mm and a lateral length of 40 mm
LaBr$_3$(Ce) plus 60 mm LaCl$_3$(Ce), all in one encapsulation.
Different DAQ systems were used for the HPGe (the
standard ISOLDE MBS system) and for the La-detectors (a CAEN digitizer DTS730
coupled to a CAEN A2818).
A pulser was added to the preamplifier of the HPGe to keep track of its
deadtime; the deadtime in the digital chain is less than in the
analogue chain.

In a two day collection campaign in May 2015 a 
total of twelve samples were collected at different
separator settings in order to cross-check for possible systematic
errors.
After sample 5 was collected extra shielding, a 10.0 mm thick Pb plate
and a 4.0 mm thick Al plate, was inserted at 10 mm distance from the
HPGe detector to further reduce its count rate.
The overall combined uncertainty in thickness is 0.5 mm.

\subsection{Production and separation of $^{11}$Be}
\label{sec:prod}
The 1.4 GeV proton beam from CERN's PS-Booster accelerator was directed onto a Ta
target (95 g in foils of 20 $\mu$m and 6 $\mu$m thickness placed in a
16 g W boat). The reaction products diffused out into a rhenium surface ion source
where Be atoms were laser ionized employing the RILIS method
\cite{Let98}. The laser settings were reoptimized on $^{9}$Be at
several occasions during the run. The ionization scheme depends on the
mass of the isotope, but not strongly: a test on $^{9}$Be with the
$^{10}$Be laser scheme gave an intensity reduction of about 35 \%.
A mass marker, containing stable Na
and Be, was used intermittently during the tuning of the separator and
target, but was turned off two hours before the first collection
started and was not used later on. The Be material in the mass marker
was provided by Alfa Aesar as a solution in 5 \% HNO$_3$.
The ions were accelerated through a
40 kV gap, the spread of ion energy out of the ion source is typically
a few eV.


The ion beam was mass separated in the High Resolution Separator
(HRS). A set of slits positioned after the HRS magnets and before the
RFQ cooler were adjusted to
delimit the beam sent into the experimental set-up. The absolute
positions of the slits were not available, but the beam full width at
half maximum was
estimated to be about 1 mm, and the final slit settings were 1.5 mm to
the high-mass side and 6.5 mm to the low-mass side (2.5 mm corresponds
to a mass difference of 0.01 u).
The masses of the isotopes and molecules that will be discussed later are
given in table \ref{tab:masses}. In the middle of the run
(between samples 8 and 9) a scan of $^{10}$Be was carried out with a 
Faraday-cup, the results are given in table \ref{tab:scan}.
They indicate that the effective width of the distribution for an
isotope is less than 0.025 u. We note for later use that the intensity
falls off by much more than two orders of magnitude when going 0.02 u
away from the nominal peak position.
The shape is rather flat at the top with
a quick fall off at the sides, which is consistent with a narrow peak
with long tails passing a slightly wider slit. (This may be compared
to the profile found for $^{11}$Be in our previous experiment, figure
1 in \cite{Rii14}.)
Note that the scan was carried out using the $^{11}$Be laser
ionization scheme, demonstrating that $^{10}$Be will be ionized with
this setting.

\begin{table}
\caption{Mass values (in u) of selected isotopes and molecules, from
  \protect\cite{Wan17}, their mass difference to $^{11}$Be and halflives.}
\label{tab:masses}       
\begin{tabular}{llll}
  \hline \noalign{\smallskip}
    & Mass (u) & Mass difference (u) & T$_{1/2}$  \\
 \noalign{\smallskip}  \hline \noalign{\smallskip}
 $^{10}$Be   & 10.01353470(9) & $-$1.0081263 & 1.39 My \\
 $^{10}$Be$^1$H & 11.02135973(9) & $-$0.0003013 & 1.39 My \\
 $^{11}$Be & 11.02166108(26) & --- & 13.8 s \\
 $^{9}$Li$^1$H$_2$ & 11.04244025(20) & 0.0207792 & 178 ms \\
 $^{11}$Li & 11.0437236(7) & 0.0220626 &  8.8 ms \\
\noalign{\smallskip}  \hline
\end{tabular}
\end{table}

\begin{table*}
\caption{Mass scan of HRS using the $^{10}$Be current reading in a
  beamline Faraday cup 
  (FC7480). Current readings below 1 pA are not reliable.}
\label{tab:scan}       
\begin{tabular}{lcccccccccc}
 \hline \noalign{\smallskip}
 mass setting (u) & 9.994 & 10.004 & 10.009 & 10.014 & 10.019 & 10.024
  & 10.029 & 10.034 & 10.039 & 10.089 \\
 current (pA) & 0.010 & 0.15  & 40 & 43 & 38 & 37 & 0.05 & 0.02 & 0.005 & $<0.001$ \\
\noalign{\smallskip}  \hline
\end{tabular}
\end{table*}

The HRS allows the distributions of $^{11}$Li and $^{11}$Be to be
separated, but tails of the neighbouring isobar may still be present.
This is a potential source of contamination since $^{10}$Be is
produced in the largest decay branch of $^{11}$Li, beta-delayed
one-neutron emission with a branching ratio about 85\%.  To further suppress the
amount of $^{11}$Li in our sample we keep the beamgate closed for 150
ms after proton impact, only a tiny fraction of the 8.75 ms $^{11}$Li nuclei
will survive this delay and will be collected.

The yield of $^{11}$Be was at the beginning of the data taking
measured in the ISOLDE tape station to $8 \cdot 10^{6}$ ions/$\mu$C proton beam.

\subsection{The collection set-up and sample change}
\label{sec:coll}
As shown in figure \ref{fig:setup},
the 40 keV ion beam passed through three collimators (all three
composed of a 1.5 mm thick Cu plate in front of an Al plate of thickness 
20 mm, 20 mm and 15 mm, respectively, they were positioned 225 mm,
80 mm and 4 mm before the collection point and had circular opening of
diameters 5 mm, 5 mm and 10 mm)
and was implanted in a small copper plate
(15$\times$20$\times$2 mm) mounted in a holder fixed to the end flange
of the beamline. The flange was taken off for sample changing.

The twelve $^{11}$Be samples and their subsequent chemical treatment
is described below.  After the final $^{11}$Be sample we collected a
source of $^{24}$Na that was used to extend the energy calibration of
the gamma detectors up to 2.75 MeV.

An attempt was made to measure the current of the ion beam on the Cu
plates, but no trustworthy readings could be obtained due to problems
in applying a reliable suppression voltage.

\subsection{The gamma-ray detection}
\label{sec:gamma}
The high-purity Ge-detector was energy and efficiency calibrated with
standard sources of $^{60}$Co, $^{133}$Ba and $^{152}$Eu placed at the
collection point. The latter
source had the best absolute intensity calibration of about 2\%, the
gamma lines at 344.3 keV, 1085.8 keV, 1089.7 keV, 1112.1 keV and 1408.0
keV were used for the efficiency calibrations, other lines were either
too weak or too close to background lines. The
$^{60}$Co source (lines at 1173.2 keV and 1332.5 keV) and the
$^{24}$Na sample (lines at 1368.6 keV and 2754.0 keV) could be used to extend the
calibration up to 2750 keV. Calibrations were done separately for the
two configurations, without and with Pb and Al plates inserted between
the calibration source and detector,
and gave final full-energy peak efficiencies at 2124 keV of
$(5.4 \pm 0.2)\cdot10^{-5}$ and $(2.7 \pm 0.2)\cdot10^{-5}$, respectively.
The calibration sources were also placed behind the collimators
closest to the collection point; the detection efficiency in the range
1100-1400 keV went down by a factor six.

\begin{figure}
  \resizebox{0.48\textwidth}{!}{%
    \includegraphics{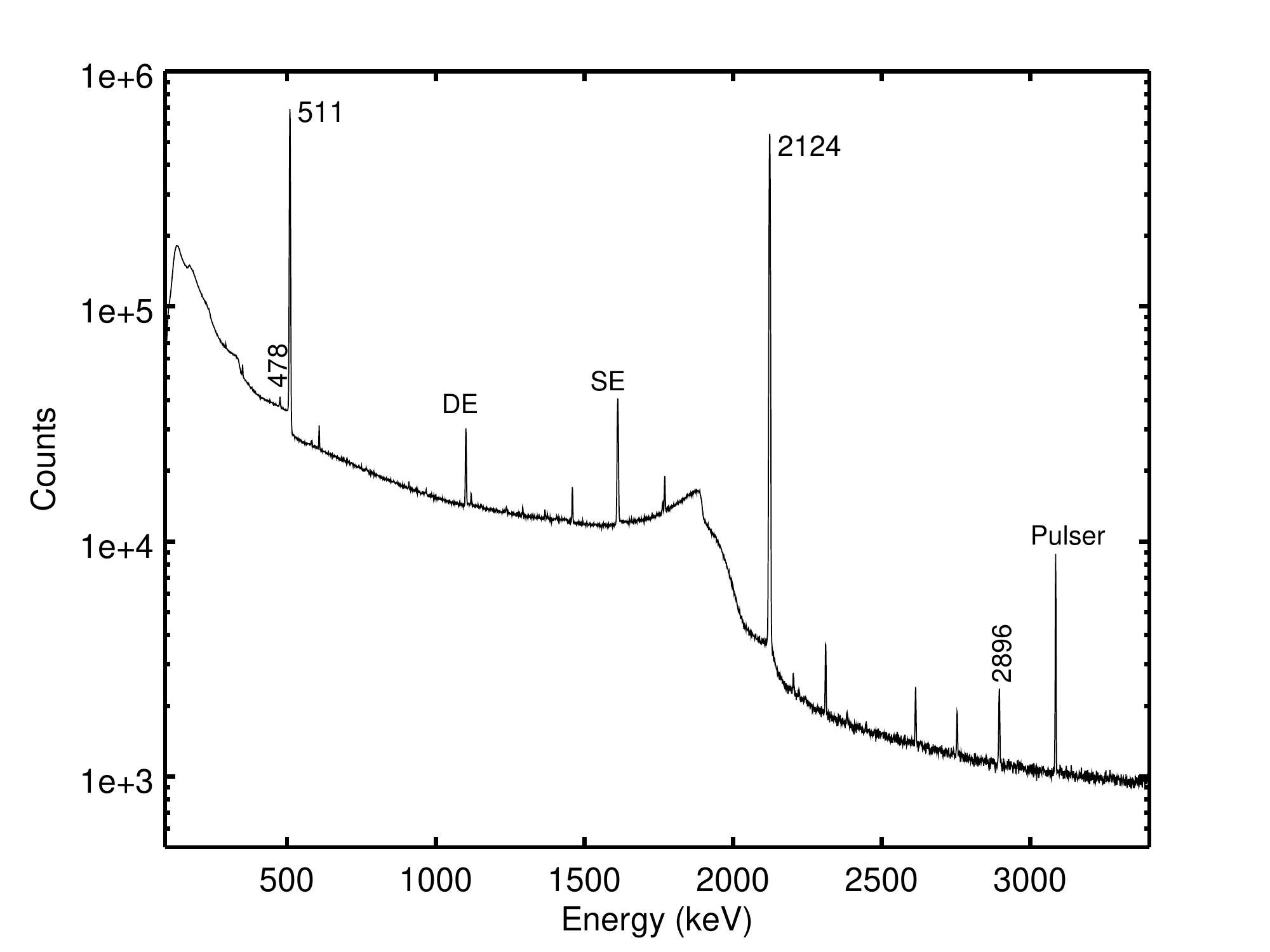}
  }
\caption{The gamma-spectrum recorded from the $^{11}$Be beam in the HPGe
  detector from sample 6. The major lines are marked by their energy
  in keV. SE and DE denote single and double escape peaks from the
  2124 keV peak. Unmarked peaks are background peaks.}
\label{fig:gammaGe} 
\end{figure}

The 2124 keV line is the dominating $\gamma$ line from $^{11}$Be. The
one at 2896 keV is also visible in our spectra, see e.g.\ figure
\ref{fig:gammaGe}, its intensity relative to the 2124 keV line is the
same for all samples where it appears and it gives a consistent yield
estimate. Several other lines at higher energy were only observed in
the LaBr detectors discussed below.  The line at 478 keV is
discussed in section \ref{sec:478keV}.

During the run we observed, apart from standard background gamma-lines
such as from $^{40}$K, gamma-lines from the decays of short-lived $^{14}$O, 
$^{24}$Na, $^{10}$C and some weaker activities, most likely also
$^{7}$Be as discussed below. The intensity of the 15 h $^{24}$Na
gamma-lines increased throughout the run, then decreased for three days
and finally increased again when the following ISOLDE run started. These
activities are therefore attributed to a background produced by
proton impact on the ISOLDE targets, either to isotopes produced
directly in the target region or produced by neutrons diffusing into
the experimental hall. We checked that no such line interferes with
the gamma-lines used in our analysis except for the 478 keV line
discussed below.

\subsubsection{Dead time}
In our previous experiment \cite{Rii14} the dead time was estimated
via the ratio of accepted to total number of triggers. As an extra
check, we here also estimate the dead time by adding a pulser whose
peak is positioned at an energy just above 3 MeV in the spectrum from
the HPGe detector.

The rate of the pulser was checked regularly during the run by
comparison to an external clock. All the tests performed were
consistent with a pulser rate of 1.001 Hz with a precision better than
$5 \cdot 10^{-4}$.  The background below the pulser peak in the gamma
spectrum was noticeable for samples with a high $^{11}$Be abundance and
was estimated from the regions to both sides of the pulser peak. The
deadtimes were deduced for each run by comparing the observed number
of events in the pulser peak with the one expected from the duration
of the run, the results are given in table \ref{tab:sample}. The
deadtimes were observed to be proportional within uncertainties to the
count rate in the gamma spectrum, the proportionality extended over
the interval from 0.1 to 5.5 kHz. The deadtime correction should be
realiable to within one percent.

\subsubsection{The 478 keV line}  \label{sec:478keV}
\begin{figure}
  \resizebox{0.48\textwidth}{!}{%
    \includegraphics{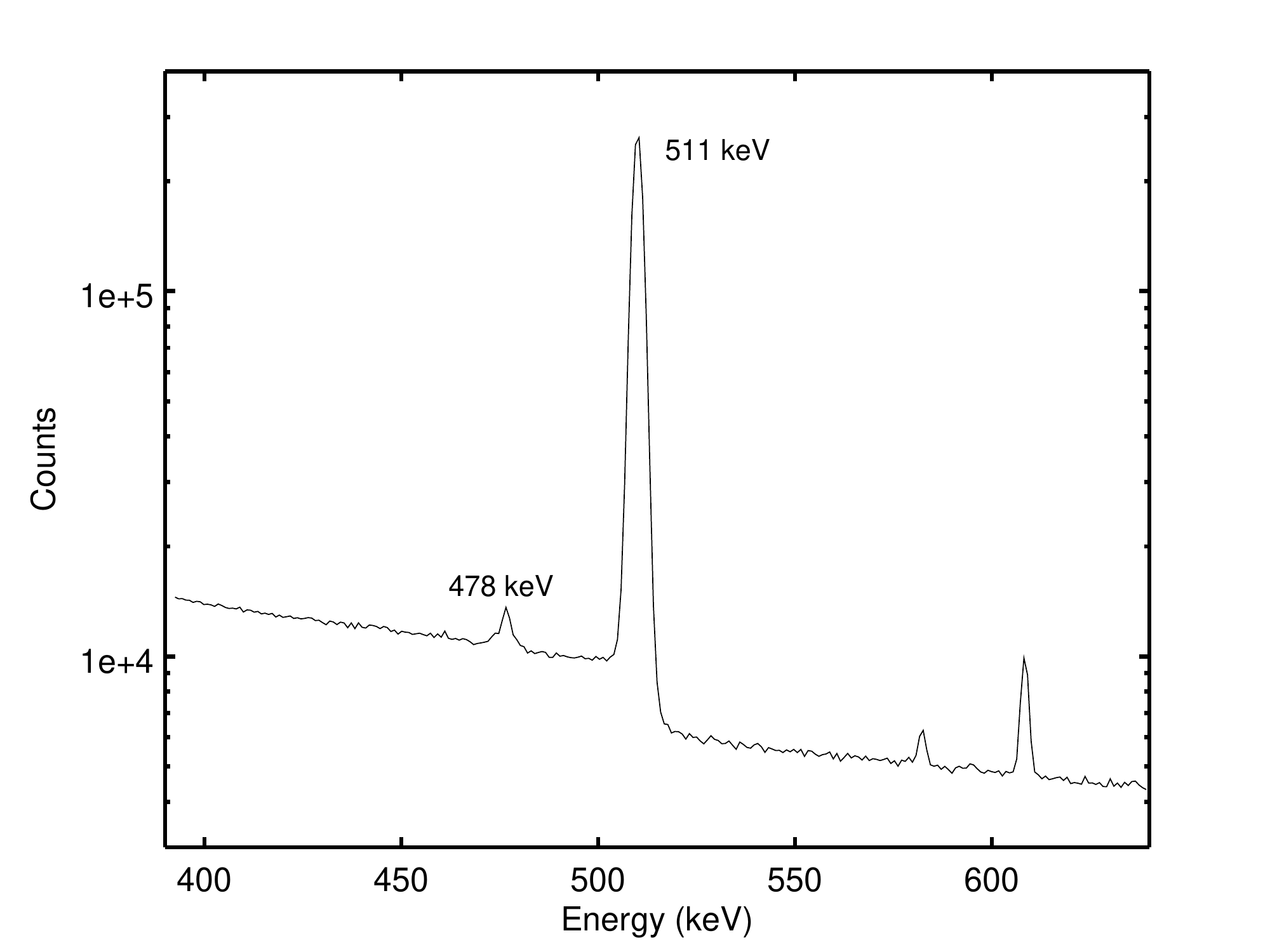}
  }
  \caption{Part of the gamma-spectrum recorded from the $^{11}$Be beam in
    the HPGe detector from sample 2. The line at 478 keV has
    contributions from a background source as well as from the
    sample, see text for details. Unmarked peaks are background peaks.}
\label{fig:gamma478} 
\end{figure}

The 478 keV line originates in beta-delayed alpha-emission to the
first excited state in $^{7}$Li. Reliable measurements of its
branching ratio
have only become available recently, the latest measurement gives
0.261(13)\% \cite{Bor13,Ref18}. The line is emitted from the moving
$^{7}$Li nucleus while it is slowed down inside the sample. The recoil
broadening gives rise to an almost triangular shape of the line. This is
clearly visible in several spectra, see figure \ref{fig:gamma478}.  In
samples with a small amount of $^{11}$Be one still observes the 478 keV
line, but now with a narrow line shape (indicative of a
non-moving source). This component is
also present in the calibration measurements and is most likely due to
a $^{7}$Be background source. Its intensity was determined to be
0.44(4) s$^{-1}$. In some samples, such as the one shown in figure
\ref{fig:gamma478}, both components are seen.

We have used the intensity of the broad component in
order to estimate whether significant parts of the $^{11}$Be activity could be
placed outside of the Cu collection plate.
It is much easier absorbed than the lines above 2 MeV, the absolute
efficiency without the extra absorbers was determined as $7.2(1) \cdot
10^{-5}$ and the intensity in samples S1, S2 and S5 agreed with the
one determined from the 2124 keV line. The extra Pb and Al absorbers
should reduce its intensity further by a factor 0.15(2), again in fair
agreement with the data.
It is therefore highly unlikely that a substantial part of the
$^{11}$Be activity is situated outside of the Cu plate, but still
visible for the HPGe detector. Such stray activity would suffer more
absorption, which would give an imbalance between the 478 keV and
2124 keV intensities.

\subsubsection{The La-detectors}
The LaBr$_3$(Ce) and CEPA4 detectors allow us to perform further
checks. The La-detectors have
much smaller volume than the HPGe detector so that absolute efficiencies
are low, they are therefore placed closer to the source collection
point. The La-detectors have as internal contaminants small activites of
$^{138}$La and $^{227}$Ac. The former gives gamma rays of energy 788.7
keV and 1435.8 keV, the latter alpha particles that give detector
signals in the range roughly from 2 to 2.5 MeV. These backgrounds are
clearly visible in background spectra, but cannot be seen when high
$^{11}$Be activities are present.

\begin{figure}
  \resizebox{0.48\textwidth}{!}{%
    \includegraphics{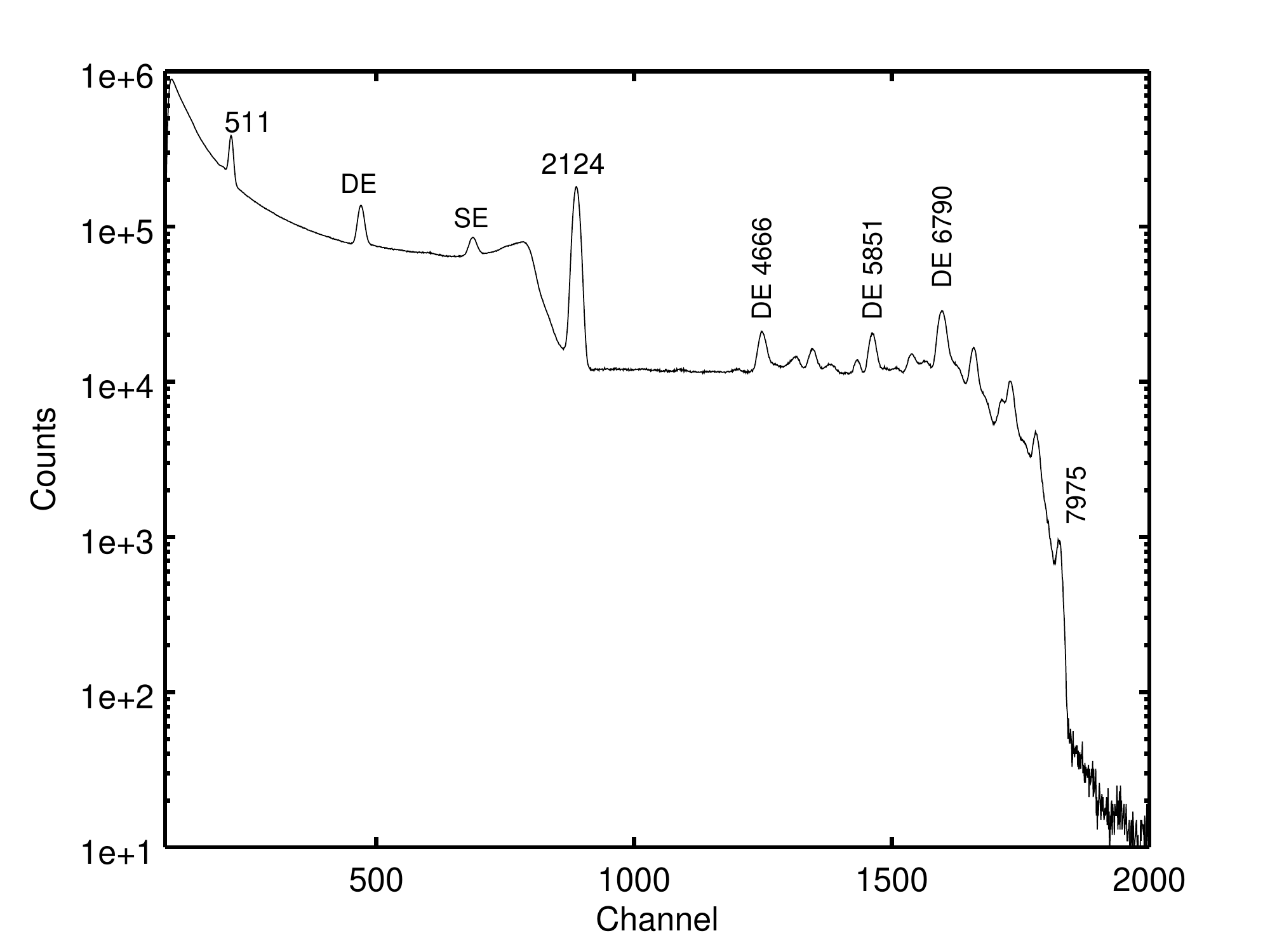}
  }
\caption{The gamma-spectrum recorded from the $^{11}$Be beam in the LaBr$_3$
  detector from sample 6. Note that the energy calibration is
  non-linear above 2 MeV. The major lines are denoted by their energy
  in keV, SE and DE denote single and double escape peaks.}
\label{fig:gammaLa} 
\end{figure}

The intrinsic efficiency of the La-detectors is good at higher gamma
energy so that the $^{11}$Be gamma lines at 4665.9 keV, 5851.5 keV,
6789.8 keV and 7974.7 keV and their escape lines are clearly seen in
the spectra. As an example the spectrum from sample 6 is
shown in figure \ref{fig:gammaLa}. The energy-channel relation is
strongly non-linear above 2 MeV, indicating that the amplification of
the photomultiplier tubes was set too
high. Due to the small volume of the LaBr$_3$ crystal
the double escape peaks are more
intense than the single escape peaks, which again are more intense
than the full energy peaks.

The energy resolution is worse than for the HPGe detector and the lines therefore more
sensitive to background components and other contamination
lines. Nevertheless, they give a valuable check of the source strength
of the samples as well as of the geometrical position of the $^{11}$Be
intensity.  The absolute detection efficiencies at 2124 keV are 
in the range of a few times $10^{-5}$ and the deduced source
intensities are consistent with the ones from the HPGe analysis.  A
further test is possible using the high-energy part of the gamma
spectrum, i.e. including all events above 3.5 MeV whether they are in
the full-energy peaks, in the escape peaks or in the Compton edge. The
background is negligble here and the number of events about double the
content of the 2124 keV full-energy peak. The variation in relative
count rates in the high-energy part and the 2124 keV peaks (in the HPGe
detector or the LaBr$_3$ detector) across the samples is less than
10\% for the LaBr$_3$ detector and less than 20 \% for the CEPA4. This
again indicates that all the visible $^{11}$Be intensity is situated
at the source collection point.

\subsection{Accelerator mass spectrometry}
\label{sec:ams}
%
%
%
The high sensitivity of accelerator mass spectrometry (AMS)
was used to deduce the number of $^{10}$Be atoms produced by
beta-delayed proton emission from the collected $^{11}$Be. To perform
the AMS measurements the $^{10}$Be had to be first chemically enriched from the
samples. To achieve this, the Cu plates were transferred to the
Helmholtz-Zentrum Dresden-Rossendorf (HZDR), where beryllium was
chemically extracted from the plates.

The first step was to partially dissolve the Cu sample in dilute nitric acid. As the
$^{11}$Be atoms impinging onto the target had an energy of only up to 40
keV the implantation depth was below 1 $\mu$m. Because of that it was not
necessary to bring the whole copper plate into solution. Instead only
the surface was dissolved, which should include all $^{10}$Be
atoms. To check for possible missed $^{10}$Be atoms and to exclude a
possible intrinsic $^{10}$Be contamination of the copper material a
second successive partial dissolution step
was performed for each Cu plate. The total weight of each copper plate
was about 5 g and the amount of copper dissolved was ranging between 10
and 30 mg for the first dissolution step and 14-36 mg for the second
dissolution step. As AMS measures isotopic ratios and not absolute
concentrations and to allow chemical separation, a well-defined amount
of stable $^{9}$Be was 
added to each solution. In this step it is very important to minimize
the amount of $^{10}$Be to be introduced together with the $^{9}$Be 
material. Commercially available $^{9}$Be material usually has too high
levels of $^{10}$Be \cite{Mer08}. About 300 $\mu$g $^{9}$Be (as a
BeCl$_{2}$ solution) extracted from a dedicated
low-level phenakite material was added to each sample. An aqueous
solution of ammonia was added to the solution, which precipitates
beryllium in the form of Be(OH)$_2$. Copper forms ammonia complexes,
which stay in the solution. After centrifugation and repeated rinsing, beryllium
hydroxide was transferred into quartz crucibles where it was first
dried on a hot plate and later ignited to beryllium oxide (BeO) at
900 $^{\circ}$C. The resulting amount of BeO from each sample was around 0.5 to
1.0 mg. The BeO was mixed with Nb powder and pressed into AMS sample
holders. Beryllium oxide samples from twelve copper plates were produced in this
way. In addition, a copper plate that was not exposed to any beam
was treated in the same way to get a “processing blank”. This allows to
check for the amount of $^{10}$Be introduced in the chemical procedure. In
total 26 AMS samples, i.e.\ 2 subsequent partial dissolutions of
twelve irradiated and one blank Cu plate, were produced.

The AMS samples were transferred to the Vienna Environmental Research
Accelerator (VERA) where the actual AMS measurement was performed in
May 2016. VERA is a dedicated AMS facility based on a 3 MV pelletron
tandem accelerator from National Electrostatics Corporation (NEC) \cite{Ste04}.

The prepared samples were introduced in a cesium sputter ion source
and a beam of BeO$^-$ was extracted. After the tandem accelerator ions
with charge state 2$^+$ were selected. To separate the $^{10}$Be ions
from the $^{10}$B background a split-anode gas ionization chamber with
a silicon nitride foil stack in front was used \cite{Ste19}.  Due to
the higher nuclear charge of B compared to Be and the resulting higher
stopping power the foil stack acts as a passive absorber reducing the
amount of $^{10}$B. The remaining boron ions entering the detector can
be identified by their different energy loss in the two regions of the
detector. The total efficiency of the AMS measurement is about
$5 \cdot 10^{-4}$. To optimize and normalize the measurement pure BeO
extracted from phenakite (machine blank) together with a standard
sample, SMD-Be-12 \cite{Akh13}, with a well-known ratio for
$^{10}$Be/$^{9}$Be of $1.704(30) \cdot 10^{-12}$ were measured together
with our samples.

\begin{table*}
\caption{For each sample, the separator setting, collection time,
  deadtime, the number of $^{11}$Be decays deduced from
  the HPGe spectra and the number of $^{10}$Be atoms deduced for the
  first and second dissolutions are listed.}
\label{tab:sample}       
\begin{tabular}{llrcccc}
 \hline \noalign{\smallskip}
  Sample  & Mass setting (u) & Time (s) & Deadtime (\%) & Number of $^{11}$Be
  & First dissolution  & Second dissolution  \\
\noalign{\smallskip}  \hline \noalign{\smallskip}
  S1  & 11.022 & 23323 & 20.5(3) & $2.41(15) \cdot 10^{11}$
  & $(180 \pm 6) \cdot 10^4$  & $(1.51 \pm 0.33) \cdot 10^4$  \\
  S2  & 11.037 & 15759 & 7.77(37) & $2.55(17) \cdot 10^{10}$
  & $(8.0 \pm 0.8) \cdot 10^4$  & $(0.74 \pm 0.28) \cdot 10^4$ \\
  S3  & 11.022 $^a$ & 8092 & 4.79(38) & $< 2.7 \cdot 10^6$
  & $(4.1 \pm 0.5) \cdot 10^4$  & $(1.22 \pm 0.31) \cdot 10^4$  \\
  S4  & 11.012 & 7646 & 4.61(31) & $1.03(7) \cdot 10^9$
  & $(6.3 \pm 0.7) \cdot 10^4$ & $(1.15 \pm 0.27) \cdot 10^4$ \\
  S5  & 11.022 & 14590 & 13.6(6) & $7.4(5) \cdot 10^{10}$
  & $(31.0 \pm 1.9) \cdot 10^4$ & $(1.08 \pm 0.32) \cdot 10^4$ \\
  S6  & 11.022 & 24048 & 13.7(5) & $2.98(27) \cdot 10^{11}$
  & $(67.0 \pm 3.4) \cdot 10^4$  & $(1.67 \pm 0.34) \cdot 10^4$  \\
  S7  & 11.044 & 7846 & 6.44(33) & $2.2(4) \cdot 10^7$
  & $(3.0 \pm 0.5) \cdot 10^4$ & $(0.80 \pm 0.32) \cdot 10^4$ \\
  S8  & 11.044 $^b$ & 7676 & 6.00(33) & $< 7.8 \cdot 10^6$
  & $(126 \pm 5) \cdot 10^4$ & $(1.37 \pm 0.30) \cdot 10^4$ \\
  S9  & 10.089 & 4290 & 6.12(42) & $< 3.3 \cdot 10^6$
  & $(1.91 \pm 0.35) \cdot 10^4$ & $(1.27 \pm 0.39) \cdot 10^4$ \\
  S10 & 10.139 $^c$ & 8553 & 5.63(30) & $< 5.5 \cdot 10^6$
  & $(1.87 \pm 0.34) \cdot 10^4$  & $(1.01 \pm 0.27) \cdot 10^4$  \\
  S11 & 11.007 & 11110 & 6.17(27) & $4.7(6) \cdot 10^7$
  & $(1.97 \pm 0.35) \cdot 10^4$ & $(1.04 \pm 0.29) \cdot 10^4$ \\
  S12 & 11.030 & 10194 & 9.97(57) & $4.8(4) \cdot 10^{10}$
  & $(8.5 \pm 0.8) \cdot 10^4$ & $(1.17 \pm 0.31) \cdot 10^4$ \\
blank $^d$  & & & & & $(1.08 \pm 0.31) \cdot 10^4$ & $(1.44 \pm 0.41) \cdot 10^4$ \\
BeO $^e$ & & & &  & $(25.6 \pm 1.2) \cdot 10^4$ &  \\
 \noalign{\smallskip} \hline \noalign{\smallskip}
\end{tabular}

$^a$ RILIS blocked; 
$^b$ $^{11}$Li run; 
$^c$ low power in RILIS;
$^d$ processing blank, low $^{10}$Be-level phenakite; 
$^e$ commercial BeO
\end{table*}

\section{Data analysis}
\label{sec:analysis}

The final deduced number of $^{11}$Be decays in each sample is
shown in table \ref{tab:sample}.  The table also lists the
collection time (uncertainty about 5 s) and HPGe detector
deadtime for each file, and the setting of the HRS mass separator. Most
runs were performed with the beamgate being closed for the first 150
ms after proton impact on target and with the RILIS ionization scheme set for
$^{11}$Be. The yield fluctuated during the whole run reaching a
maximum of around $10^7$ $^{11}$Be/s. The fluctuations were partly due
to slow drifts in RILIS so the lasers were monitored and reoptimized
when needed. In particular, at the
end of collection of sample S10 the RILIS power was found to be very low so that no
conclusions can be drawn from that sample. Sample S3 was on purpose taken
with RILIS being off and sample S8 (on the mass of $^{11}$Li) was also
taken with RILIS off and with an ``inverted'' beamgate, i.e.\ only
open in the first 150 ms after proton impact.

The $^{9}$Li$^1$H$_2^+$ ion is stable \cite{San05} and may, if
produced in our target and ion source, contribute to the beta count
rate close to the $^{11}$Li mass position. However, it is unlikely to
be a dominating component and it will in any case not contribute to
the production of $^{10}$Be.

%

The AMS analysis results 
are given in the last two columns of table \ref{tab:sample}. 
The amount of $^{10}$Be found in the second dissolutions are in the
range of (1--1.5) $ \cdot 10^4$ atoms. In the first dissolution only
the processing blank
sample reaches this low value, whereas values around $1.9 \cdot 10^4$
atoms are found for samples S9, S10 and S11.
The maximum amount of $^{10}$Be found in the first dissolutions is two
orders of magnitude higher.

Note that the number of $^{10}$Be atoms observed for sample S9
converts into an upper limit of about 6 atoms/s for the $^{10}$Be
intensity at the mass setting 10.089 u. This is orders of magnitude below the
upper limit on the
current recorded for the same mass setting in table \ref{tab:scan} 
where the maximum intensity, at the peak position, converts to $2.7 \cdot
10^8$ ions/s. This strongly indicates that the ``high mass tail'' of
the $^{10}$Be distribution, which corresponds to ions that have higher
kinetic energy than the average or have been scattered on the path
through the magnets, will have negligible intensity at mass 11.
We note that this high level of purity would be difficult to obtain
with a one stage magnetic separator.

\section{Discussion of results}
\label{sec:res}

The observed line profile for $^{11}$Be as the separator mass setting
is increased is similar to the one seen in table \ref{tab:scan} for
$^{10}$Be, there is a steeper fall-off towards low mass settings
(samples S11 and S4, in contrast to samples S12 and S2)
consistent with the asymmetry of the slit settings. The different
samples allow a systematic check of the three potentially worrying
backgrounds that were also identified in our earlier work \cite{Bor13,Rii14}.

\emph{The first} is direct production of $^{10}$Be being scattered to the
wrong mass value somewhere in the separator. Our two steps in the
magnetic separation and the electrostatic transport makes such a
background rather unlikely, but the $^{10}$Be peak is more intense than
$^{11}$Be. It is therefore crucial that samples S9 and S11 have very
little $^{10}$Be content.  As argued in the previous section this
rules out this background source.

\emph{The second} possibility is direct production of $^{11}$Li whose largest
decay branch produces $^{10}$Be. We do indeed see clear $^{10}$Be
production in sample S8 where we explicitly run for collecting
$^{11}$Li. The intensity corresponds to collecting 190(8) $^{11}$Li/s,
which is a reasonable yield at these running conditions for the
target.  The corresponding gamma lines from the $^{11}$Li decay were
too weak to be seen, as would be expected. The crucial point is 
that for sample S7 at our standard running conditions very little of
the $^{11}$Li leaked through, at most 3\%. This amount is further reduced by
more than two orders of magnitude when moving 0.02 u down (as
noted above), so we can safely conclude that the
possible contamination from $^{11}$Li will be much less than the
signal observed on the $^{11}$Be mass.

\emph{The third} possibility is ionization without break-up of neutral
$^{10}$Be$^1$H molecules,
a signal that would appear directly on the $^{11}$Be
position. We have earlier \cite{Bor13,Rii14} argued based on the known
properties of this molecule \cite{Bub07} that it is highly unlikely
that laser ionization gives a molecular ion rather than breaking up the
molecule, in particular in the strong laser fields in the source. If this
is correct molecular ions are surface ionized and their yield should be at
least the same when the lasers are turned off, i.e.\ the conditions
for sample S3. We see very little signal there, which excludes
surface ionization of neutral BeH molecules in the ion source as
significant background to the signal at mass 11.022 u.
If the small BeH signal from sample S3 is indicative of a non-zero
contribution it would be peaked at the $^{10}$Be$^1$H mass and should
therefore at other mass values decrease in intensity in a similar way
that the $^{11}$Be intensity does.

Turning now to the deduced branching ratios for beta-delayed
proton emission, these are collected in table \ref{tab:res} and
divided into samples taken on the central mass value and those taken
sligthly off mass, the latter ordered after the observed number of
$^{11}$Be. To evaluate the branching ratios a background of $1.9 \cdot
10^4$ atoms was subtracted from the number of $^{10}$Be atoms.
(The sensitivity limit of our set-up would therefore, for collections
of duration up to a day, correspond to branching ratios of order $10^{-7}$.)
The result for sample S4 is surprisingly high,
but we note that it corresponds to the
lowest recorded numbers of $^{10}$Be and $^{11}$Be out of the six samples shown.
For the other five samples, the overall order of magnitude is similar
to that obtained earlier, namely the value of $2.5(2.5) \cdot
10^{-6}$ from \cite{Bor13} and the value of $8.3(9) \cdot 10^{-6}$
from \cite{Rii14}, but the new values do clearly not agree internally.

\begin{table}
\caption{Branching ratios for beta-delayed proton decay of $^{11}$Be
  derived from the values in table \protect\ref{tab:sample}.}
\label{tab:res}       
\begin{tabular}{lccc}
 \hline \noalign{\smallskip}
Sample, mass centred & S1 & S5 & S6 \\
Branch ($10^{-6}$) & 7.4(5) & 3.9(4) & 2.2(2) \\  
 \noalign{\smallskip} \hline  \noalign{\smallskip}
Sample, off centre & S12 & S2 & S4 \\
Branch ($10^{-6}$) & 1.4(2) & 2.4(4) & 43(8) \\  
 \noalign{\smallskip} \hline
\end{tabular}
\end{table}

\subsection{Conjecture}
\label{sec:conjec}
One possible interpretation of the disagreeing branching ratios
could be the presence of a source of $^{10}$Be that is displaced
slightly below the $^{11}$Be mass values and decreases in intensity
with time. A gradual $^{10}$Be intensity decrease could explain the
decreasing branching ratios for the
samples on the central mass value: S1, S5 and S6. The displacement of
the source would be needed to explain why S2 is low (and similar for
S12) and, in particular, why S4 is much higher than the other values:
From sample S4 to sample S5 the $^{11}$Be intensity per time increases
a factor 38(3), whereas the increase for $^{10}$Be is an order of
magnitude less, a factor 3.5(6). From sample S1 to sample S2 the
corresponding intensities per time decrease by factors of 6.4(6) and
20(3) for $^{11}$Be and $^{10}$Be (in this case part of the change
could be due to a drop in the target yield since the deadtime was
observed to decrease slightly during collection of sample S1). This
pattern suggests that the $^{10}$Be intensity peaks at lower mass
values than the $^{11}$Be intensity.

A simple estimate can be made by assuming Gaussian intensity profiles,
using the mass difference of 0.0003 u between $^{10}$Be$^1$H and
$^{11}$Be from table \ref{tab:masses} and a width parameter of 0.0017
u derived from the beam profile estimate in section
\ref{sec:prod}. This gives the correct trend, but with too small
intensity differences for the two components. The beam profile will be
more complex than a Gaussian, but this nevertheless may point towards
a larger effective mass difference than 0.0003 u.

The following hypothetical mechanism might explain our
observations. We stress that we do not have any direct evidence for
it, but present it as the only coherent --- and possibly plausible ---
explanation we have been able to find for our data.

The suggestion is that $^{10}$Be$^1$H$^+$ is formed from ionized
$^{10}$Be$^+$ reacting with hydrogen (or water vapour) during the
brief interval during which the ion is extracted from the ion
source. ISOLDE targets contain hydrogen in small amounts and it is
pumped out rather slowly when targets are put under vacuum. The
gradual reduction of the amount of available hydrogen could then
explain the decreasing size of our signal.
The modelling of the chemical reaction inside (or on the way out of)
the ion source will be complex. Our laser setting did ionize $^{10}$Be
so we are looking at processes with relative probability of order
$10^{-7}$. However, some supporting evidence comes from the following
observations of molecular beams in related environments:

At ISOLTRAP an experiment on $^{79}$Cu$^+$ gave evidence for the
molecule $^{65}$Cu$^{12}$CH$_2^+$ in a MR-TOF spectrum \cite{Wel17}.
A resonance ionization mass spectrometry experiment at the University of
Mainz identified a background of $^{40}$CaH$^+$ ions produced by
collisions of $^{40}$Ca$^+$ with hydrogenous residual gas molecules \cite{Mul01},
the experimental conditions giving a relative intensity of $2 \cdot 10^{-5}$
for the hydride.  At LISOL \cite{Kud03} (and other places) detailed
studies of the interaction of ions with buffer gas- and impurity
molecules led to observations of e.g.\ Co(H$_{2}$O)$^+$.

The suggested mechanism would also be expected to be present at the
earlier beamtimes. The large signal reported in \cite{Rii14} was from
a sample taken shortly after the target had been put on-line and may
therefore be due to the same effect. The smaller signal reported in
\cite{Bor13} could be due to different purity of the target, e.g.\ a
sample taken at a later time. A similar explanation may apply for the
limit of $2 \cdot 10^{-6}$ of $^{11}$Be on mass $^{12}$Be quoted in
\cite{Rii14}. 

Two obvious future checks of this mechanism are to look for $^{12}$Be
on mass $^{13}$Be and for $^{7}$Be on mass $^{8}$Be. If the mechanism
can be confirmed there are several ways to reduce its effect. One may
run with a narrow-band laser setting for $^{11}$Be to reduce the
simultaneous ionization of $^{10}$Be, or one may
attempt to dissociate the BeH molecules, one possibility being to
pass the ion beam through a thin C-foil (or a gas target).

\section{Summary and outlook}
\label{sec:outlook}
The aim of this experiment was to check our previous claim
\cite{Rii14} of a surprisingly large branching ratio of
$8.3(9) \cdot 10^{-6}$ for beta-delayed proton decay of $^{11}$Be.
The variation of separator settings for the different collected
samples has allowed to exclude all previously considered sources of
contamination.  Still, the current results for the branching ratio are
inconclusive and not consistent with the previous value. We conjecture
that a better understanding of molucular formation in the ion source
is needed, and that there is a clear risk that our signal stems from
this mechanism.

At the current level of understanding the most likely interpretation
of the branching ratio results collected in table \ref{tab:res}
is that a background source dominates all values execpt possibly the
one from sample S12. A conservative 95\% confidence level upper limit
of $2.2 \cdot 10^{-6}$ for the branching ratio for beta-delayed proton
decay can be obtained from sample S12 from the ratio of the number of
$^{10}$Be and $^{11}$Be atoms without any background subtraction.
We note that if neutron dark decays are responsible for the neutron
decay anomaly \cite{For18}, a sizable part of the energy range for the
new ``dark'' particles is excluded by our limit \cite{Pfu18}.


We outlined above possible ways to improve upon the present procedure
for measuring the decay, but of course encourage experiments with
alternative procedures, preferably ones that detect the proton. Due to
the low branching ratio such experiments are rather difficult, but the
recent interest in this decay has prompted several groups to attempt
detection inside a TPC either by optical detection at ISOLDE
\cite{Cie16} or electrical detection at TRIUMF \cite{Ayy19}. As the
main background in such detectors could be the beta-delayed alpha
branch such experiments should be sensitive down to an interesting
region of branching ratios. The recent report from the TRIUMF
experiment \cite{Ayy19} of a positive identification of the
beta-delayed proton decay with a branching ratio of
$13(3) \cdot 10^{-6}$ and a narrow proton energy distribution peaking
around 178(20) keV is very interesting.
This value does not agree with our upper limit. However, one possible
explanation may be that the TRIUMF experiment observed (at least
partially) the beta-delayed triton branch, since protons and tritons
would be more difficult to distinguish in their set-up. Our results
only address the p+$^{10}$Be channel.




\begin{acknowledgement}
We would like to thank the ISOLDE technical teams for their efforts
and T. Andersen, J. Heinemeier, B. Marsh, H. Ravn, S. Rothe, P. Van Duppen and
K. Wendt for discussions. 
We would like to acknowledge support from the European Union's Horizon
2020 research and innovation programme under grant agreement
no. 654002, from the Spanish research projects
FPA2015-64969-P, FPA2015-65035-P and
FPA2017-87568-P (MINECO/FEDER, UE), and from
the Danish Council for Independent Research (DFF - 4181-00218).
\end{acknowledgement}


\begin{thebibliography}{999}
\bibitem{Pfu12} M. Pf\"{u}tzner, M. Karny, L.V. Grigorenko and K. Riisager,
  Rev. Mod. Phys. \textbf{84}, (2012) 567.
\bibitem{Bla08}B. Blank and M.J.G. Borge,
  Prog. Part. Nucl. Phys. \textbf{60}, (2008) 403.
\bibitem{Jon01} B. Jonson and K. Riisager, Nucl. Phys. \textbf{A693},
  (2001) 77.
\bibitem{Bay11} D. Baye and E.M. Tursonov, Phys. Lett. B \textbf{696}, (2011) 464.
\bibitem{Ang98} J.C. Ang\'{e}lique et al., IS374 experiment proposal,
  CERN/ISC 98-6 ISC/P99 (1998).
\bibitem{Bor13} M.J.G. Borge, L.M. Fraile, H.O.U. Fynbo, B. Jonson,
  O.S. Kirsebom, T. Nilsson, G. Nyman, G. Possnert, K. Riisager and
  O. Tengblad, J. Phys. G \textbf{40}, (2013) 035109. 
\bibitem{Rii14} K. Riisager, O. Forstner, M.J.G. Borge, J.A. Briz,
  M. Carmona-Gallardo, L.M. Fraile, H.O.U. Fynbo, T. Giles,
  A. Gottberg, A. Heinz, J.G. Johansen, B. Jonson, J. Kurcewicz,
  M.V. Lund, T. Nilsson, G. Nyman, E. Rapisarda, P. Steier,
  O. Tengblad, R. Thies and S.R. Winkler et al., Phys. Lett. B \textbf{732}, (2014) 305.
\bibitem{For18} B. Fornal and B. Grinstein, Phys. Rev. Lett. \textbf{120}, (2018) 191801.
\bibitem{Pfu18} M. Pf\"{u}tzner and K. Riisager, Phys. Rev. \textbf{C97}, (2018) 042501.
\bibitem{Ref18} J. Refsgaard, J. B\"{u}scher, A. Arokiaraj, H.O.U. Fynbo, R. Raabe,
  and K. Riisager, Phys. Rev. \textbf{C99} (2019) 044316.
\bibitem{Nac15}E. N\'{a}cher, M. M{\aa}rtensson, O. Tengblad,
  H. \'{A}lvarez-Pol, M. Bendel, D. Cortina-Gil, R. Gernh\"{a}user,
  T. Le Bleis, A. Maj, T. Nilsson, A. Perea, B. Pietras, G. Ribeiro,
  J. S\'{a}nchez del R\'{\i}o, J. S\'{a}nchez Rosado, A. Heinz, B. Szpak,
  M. Winkel, M. Zieblinski, Nucl. Instr. Meth. \textbf{A769}, (2015) 105.
\bibitem{Let98} J. Lettry, R. Catherall, G. J. Focker, O. C. Jonsson,
  E. Kugler, H. Ravn, C. Tamburella, V. Fedoseyev, V. I. Mishin,
  G. Huber, V. Sebastian, M. Koizumi, and U. K\"{o}ster,
  Rev. Sci. Instr. \textbf{69}, (1998) 761. 
\bibitem{Wan17} Meng Wang, G. Audi, F.G. Kondev, W.J. Huang, S. Naimi,
  Xing Xu, Chinese Physics \textbf{C41}, (2017) 030003.
\bibitem{Mer08} S. Merchel et al., Nucl. Instr. Meth. \textbf{B266}, (2008) 4921.
\bibitem{Ste04} P. Steier et al., Nucl. Instr. Meth. \textbf{B223},
  (2004) 67.
\bibitem{Ste19} P. Steier et al., Int. J. Mass Spectrometry \textbf{444},
  (2019) 116175.
\bibitem{Akh13} S. Akhmadeliev, R. Heller, D. Hanf, G. Rugel,
  S. Merchel, Nucl. Instr. Meth. \textbf{B294}, (2013) 5.
\bibitem{San05} C. Sanz, E. Bodo and F.A. Gianturco, Chemical Physics
  \textbf{314}, (2005) 135.
\bibitem{Bub07} S. Bubin and L. Adamowicz, J.Chem.Phys. \textbf{126}, (2007) 214305.
\bibitem{Wel17} A. Welker et al., Phys. Rev. Lett. \textbf{119},
  (2017) 192502.
\bibitem{Mul01} P. M\"{u}ller et al., Fresenius
  J. Anal. Chem. \textbf{370}, (2001) 508.
\bibitem{Kud03} Yu. Kudryavtsev et al., Nucl. Instr. Meth. \textbf{B204}, (2003) 336.
\bibitem{Cie16} A.A. Ciemny et al., CERN-INTC-2016-048/INTC-P-479.
\bibitem{Ayy19} Y. Ayyad et al., Phys. Rev. Lett. \textbf{123}, (2019) 082501.
 
\end{thebibliography}
\end{document}